\documentclass[9pt,twocolumn,twoside]{pnas-new}

\templatetype{pnasresearcharticle} 

\usepackage{graphicx}
\usepackage{amssymb, amsfonts, amsmath}
\usepackage{here}
\usepackage{bm}
\usepackage{color}
\usepackage{here}
\usepackage{xr}

\usepackage{textcomp}

\makeatletter

\newlength{\figwidth}
 \newcommand{\FigCap}[1]{\uppercase{#1}}	

\makeatother

	\title{
	{Two superconducting states with broken time-reversal symmetry in FeSe$\bm{_{1-x}}$S$\bm{_{x}}$}
} 

\author[a, b]{Kohei Matsuura}
\author[a]{Masaki Roppongi}
\author[a]{Mingwei Qiu}
\author[c]{Qi Sheng}
\author[d, e]{Yipeng Cai}
\author[c]{Kohtaro Yamakawa}
\author[c]{Zurab Guguchia}
\author[d, e]{Ryan P. Day}
\author[f, d]{Kenji M. Kojima}
\author[d, e]{Andrea Damascelli}
\author[a]{Yuichi Sugimura}
\author[a]{Mikihiko Saito}
\author[a]{Takaaki Takenaka}
\author[a]{Kota Ishihara}
\author[a,g]{Yuta Mizukami}
\author[a]{Kenichiro Hashimoto}
\author[h]{Yilun Gu}
\author[h]{Shengli Guo}
\author[h]{Licheng Fu}
\author[h]{Zheneng Zhang}
\author[h]{Fanlong Ning}
\author[i]{Guoqiang Zhao}
\author[i]{Guangyang Dai}
\author[i]{Changqing Jin}
\author[j]{James W. Beare}
\author[j, f]{Graeme M. Luke}
\author[c,1]{Yasutomo J. Uemura}
\author[a,1]{Takasada Shibauchi}

\affil[a]{Department of Advanced Materials Science, University of Tokyo, Kashiwa,
Chiba 277-8561, Japan}
\affil[b]{Present address: Department of Applied Physics, University of Tokyo, Bunkyo-ku, Tokyo 113-8656, Japan}
\affil[c]{Department of Physics, Columbia University, New York, NY 10027 USA}
\affil[d]{Quantum Matter Institute, University of British Columbia, Vancouver, BC V6T 1Z4, Canada}
\affil[e]{Department of Physics \& Astronomy, University of British Columbia, Vancouver, BC V6T 1Z1, Canada}
\affil[f]{Centre for Molecular and Materials Science, TRIUMF, Vancouver BC V6T 2A3, Canada}
\affil[g]{Present address: Department of Physics, Tohoku University, Aoba-ku, Sendai, Miyagi 980-8578, Japan}
\affil[h]{Department of Physics, Zhejiang University, Hangzhou 310027, China}
\affil[i]{Beijing National Laboratory for Condensed Matter Physics; Institute of Physics,
	Chinese Academy of Sciences; School of Physics,
	University of Chinese Academy of Sciences, Beijing 100190, China}
\affil[j]{Department of Physics and Astronomy, McMaster University, Hamilton, ON L8S 4M1 Canada}

\leadauthor{Matsuura} 

\significancestatement{
	When metal becomes superconducting, the Fermi surface becomes unstable, and a superconducting gap forms. In unconventional superconductors, the gap varies in momentum space and can become zero at points or lines on the original Fermi surface. More exotic superconducting states with time-reversal symmetry breaking (TRSB) have been theoretically proposed, but such states are rarely found. Our muon experiments on iron chalcogenide superconductors FeSe$\bm{_{1-x}}$S$\bm{_{x}}$ in the orthorhombic and tetragonal phases provide evidence for TRSB below the transition temperatures. Moreover, the TRSB superconducting state in the tetragonal phase has a unique feature that not all electrons become superconducting. Our results reveal that this system exhibits two distinct superconducting states both with TRSB, which provides new research directions for exotic superconductivity.

}

\authorcontributions{
	T.S. and Y.J.U. conceived and supervised the project. 
	K.M., M.Q., Q.S., K.Y., Y.S., and M.S. synthesized the single crystals. 
	K.M., M.Q., Q.S., K.Y., Z.G., Y.P.C., R.P.D., K.M.K., A.D., T.T., Y.G., S.L.G., L.C.F., Z.Z., F.L.N., G.Q.Z., G.Y.D., C.Q.J., J.W.B., G.M.L., and Y.J.U. performed $\mu$SR measurements. K.M., M.Q., K.I., Y.M., K.H., Y.J.U., and T.S. analyzed the superfluid density data.
	All authors discussed the results. K.M. and T.S. wrote the paper with inputs from all authors. }
\authordeclaration{The authors declare no conflict of interest.}
\correspondingauthor{\textsuperscript{1}To whom correspondence should be addressed. E-mail: yu2@columbia.edu or shibauchi@k.u-tokyo.ac.jp}


\begin{abstract}
Iron-chalcogenide superconductors FeSe$\bm{_{1-x}}$S$\bm{_{x}}$ possess unique electronic properties such as non-magnetic nematic order and its quantum critical point. The nature of superconductivity with such nematicity is important for understanding the mechanism of unconventional superconductivity. A recent theory suggested the possible emergence of a fundamentally new class of superconductivity with the so-called Bogoliubov Fermi surfaces (BFSs) in this system. However, such an {\em ultranodal} pair state requires broken time-reversal symmetry (TRS) in the superconducting state, which has not been observed experimentally. Here we report muon spin relaxation ($\bm{\mu}$SR) measurements in FeSe$\bm{_{1-x}}$S$\bm{_{x}}$ superconductors for $\bm{0\le x \le 0.22}$ covering both orthorhombic (nematic) and tetragonal phases. We find that the zero-field muon relaxation rate is enhanced below the superconducting transition temperature $\bm{T_{\rm c}}$ for all compositions, indicating that the superconducting state breaks TRS both in the nematic and tetragonal phases. Moreover, the transverse-field $\bm{\mu}$SR measurements reveal that the superfluid density shows an unexpected and substantial reduction in the tetragonal phase ($\bm{x>0.17}$). This implies that a significant fraction of electrons remain unpaired in the zero-temperature limit, which cannot be explained by the known unconventional superconducting states with point or line nodes. The time-reversal symmetry breaking and the suppressed superfluid density in the tetragonal phase, together with the reported enhanced zero-energy excitations, are consistent with the ultranodal pair state with BFSs. The present results reveal two different superconducting states with broken TRS separated by the nematic critical point in FeSe$\bm{_{1-x}}$S$\bm{_{x}}$, which calls for the theory of microscopic origins that account for the relation between the nematicity and superconductivity.
\end{abstract}

\dates{This manuscript was compiled on \today}
\doi{\url{www.pnas.org/cgi/doi/10.1073/pnas.XXXXXXXXXX}}

\begin{document}

\maketitle
\thispagestyle{firststyle}
\ifthenelse{\boolean{shortarticle}}{\ifthenelse{\boolean{singlecolumn}}{\abscontentformatted}{\abscontent}}{}


Recent studies of FeSe$_{1-x}$S$_{x}$ using high-quality single crystals up to $x\sim 0.25$ have shown that electronic nematic order sets in at the structural transition temperature $T_{\rm s}$, which can be completely suppressed by S substitution without inducing any magnetic ordering under ambient pressure~\cite{Shibauchi2020,Coldea2021}. The phase diagram of FeSe$_{1-x}$S$_{x}$ exhibits a nematic quantum critical point at $x\approx0.17$~\cite{Hosoi2016,Ishida2022}, near which anomalous non-Fermi liquid transport properties have been reported~\cite{Licciardello2019,Huang2020}. The superconducting gap structure of FeSe$_{1-x}$S$_{x}$ inside the nematic (orthorhombic) phase is very anisotropic, and it has been discussed that $\Delta(\bm{k})$ exhibits accidental line nodes or deep minima~\cite{Kasahara2014,Watashige2015,Xu2016,Sprau2017,Hashimoto2018,Liu2018}, indicating the unconventional nature of its superconductivity. 
In the nematic phase with $C_2$ symmetry, the superconducting gap has two-fold rotational symmetry, which can be described by the sum of two contributions $\Delta_s$ and $\Delta_d$ having $s$-wave and $d$-wave symmetries, respectively~\cite{Sigrist1996,Kang2018}. We should also consider the effect of nematic domains whose orientations are orthogonal to each other, and in neighboring domains, one of the $s$- and $d$-wave contributions has opposite signs, such as $\Delta_s\pm\Delta_d$~\cite{Sigrist1996}. Scanning tunneling microscopy (STM) and angle-resolved photoemission spectroscopy (ARPES) studies in FeSe have shown that the nematic twin boundary affects the low-energy excitation spectrum, opening a gap near the boundary~\cite{Watashige2015,Hashimoto2018}. It has been suggested that across the twin boundary, $\Delta_s+\Delta_d$ may be transformed into $\Delta_s-\Delta_d$ through a TRS breaking (TRSB) state $\Delta_s+e^{{\rm i}\theta}\Delta_d$ with $0<\theta<\pi$, which is consistent with the gap opening near the boundary. However, direct evidence of TRSB has not been reported in superconducting FeSe. 

In the tetragonal phase of FeSe$_{1-x}$S$_{x}$ ($x>0.17$), the nematic order disappears and thus no twin boundary is observed in the STM measurements~\cite{Hanaguri2018}. It has been reported that although the electronic structure evolves smoothly across the critical concentration~\cite{Hanaguri2018,Coldea2019}, the superconducting properties change abruptly~\cite{Sato2018,Hanaguri2018,Mizukami2021}. In the tetragonal phase, the STM conductance spectra in the superconducting state have found a substantial zero-bias value, and the specific heat and thermal conductivity show large residuals at low temperatures. 

\begin{figure*}
	\centering
	\includegraphics[width=1\textwidth]{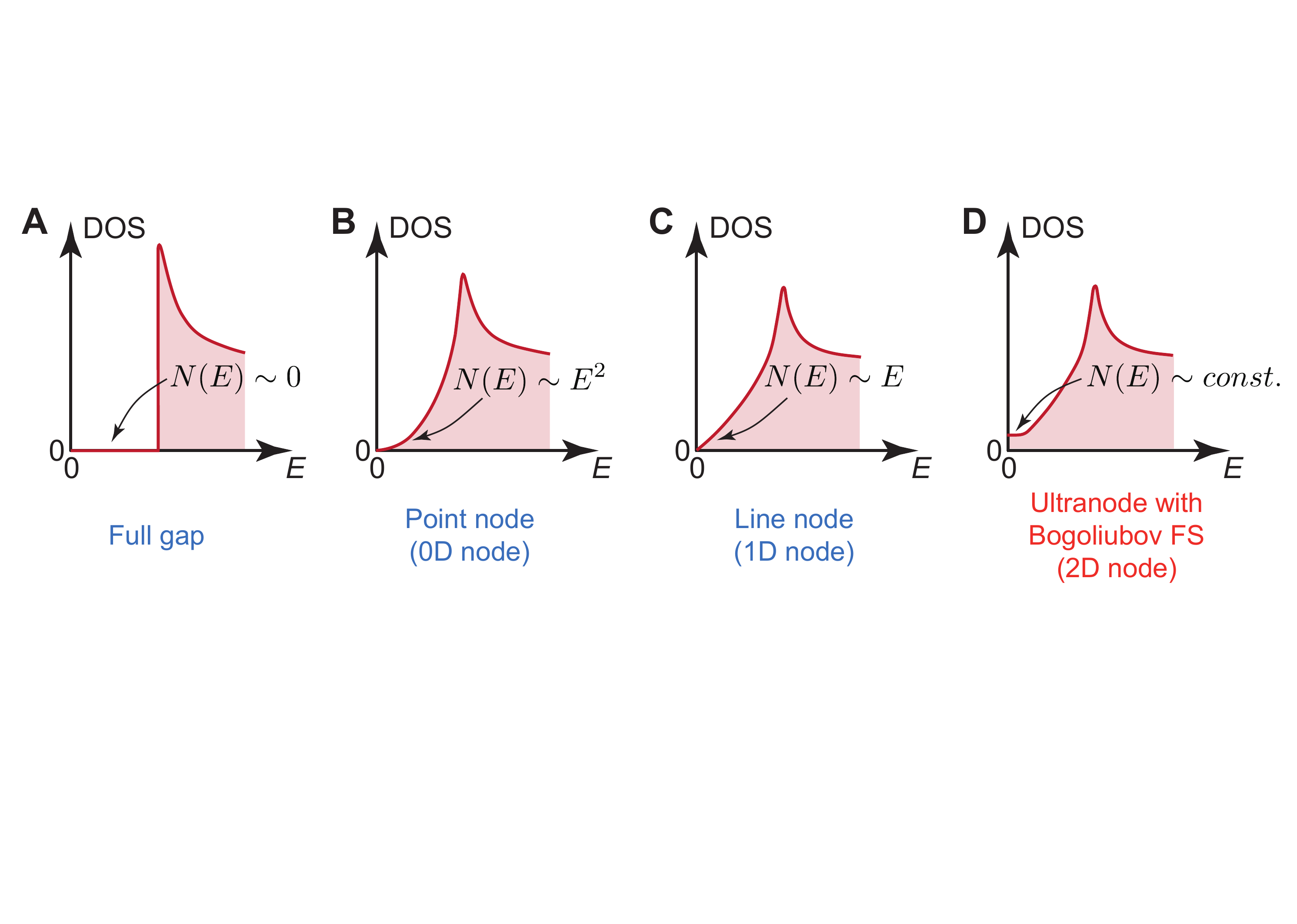}
	\vspace{-15mm}
	\caption{Classification of low-energy quasiparticle excitations with different nodal structure of superconducting gap. (A) Systematic energy dependence of quasiparticle density of states $N(E)$ for fully gapped superconductors with no nodes. (B) For superconductors with point nodes, the Fermi surfaces in the normal state are gapped except for the nodal points and $N(E)$ follows $E^2$ dependence at low energies. (C) When the superconducting gap has lines of nodes, $N(E)$ obeys $E$-linear dependence at low energies in the absence of magnetic field. (D) A new class of superconducting gap structure exhibits 2D surfaces of nodes in the superconducting state. In this case, TRS has to be broken, and extended zero-energy excitations give finite $N(0)$, similar to metallic states with Fermi surfaces. 
	}
	\label{fig:class}
\end{figure*}

Recent theoretical studies have suggested that FeSe$_{1-x}$S$_{x}$ may have an ultranodal superconducting state with BFSs, which can account for these unusual properties reported in the tetragonal phase.
In conventional $s$-wave superconductors the gap is almost isotropic in momentum space, and the quasiparticle density of state $N(E)$ is essentially absent at low energies (Fig.\:\ref{fig:class}\FigCap{a}). In unconventional superconducting states, the superconducting gap $\Delta(\bm{k})$ on the normal-state Fermi surfaces may have points or lines of zeros (nodes). Consequently, the low-energy quasiparticle excitations are governed by the nodal structure, as summarized in Fig.\:\ref{fig:class}\FigCap{b},\FigCap{c}. In contrast, the novel ultranodal superconducting state has extended two-dimensional surfaces of nodes ($\Delta(\bm{k})=0$). These BFSs lead to extended zero energy states at zero field like a normal metal (Fig.\:\ref{fig:class}\FigCap{d}). It has been shown theoretically by Agterberg {\it et al.}~\cite{Agterberg2017,Brydon2018} that such an exotic pairing state with BFSs can be energetically stable under certain conditions, if the system breaks TRS. A recent theoretical study considers a multiband system with intraband and interband interactions with broken TRS, and suggests an ultranodal pair state with BFSs, which may apply to FeSe-based superconductors~\cite{Setty2020,Setty2020a}.
It is then extremely important to clarify whether TRS is broken or not in the FeSe$_{1-x}$S$_{x}$ experimentally.

\section*{RESULTS}
\begin{figure*}
	\centering
	\includegraphics[width=1\textwidth]{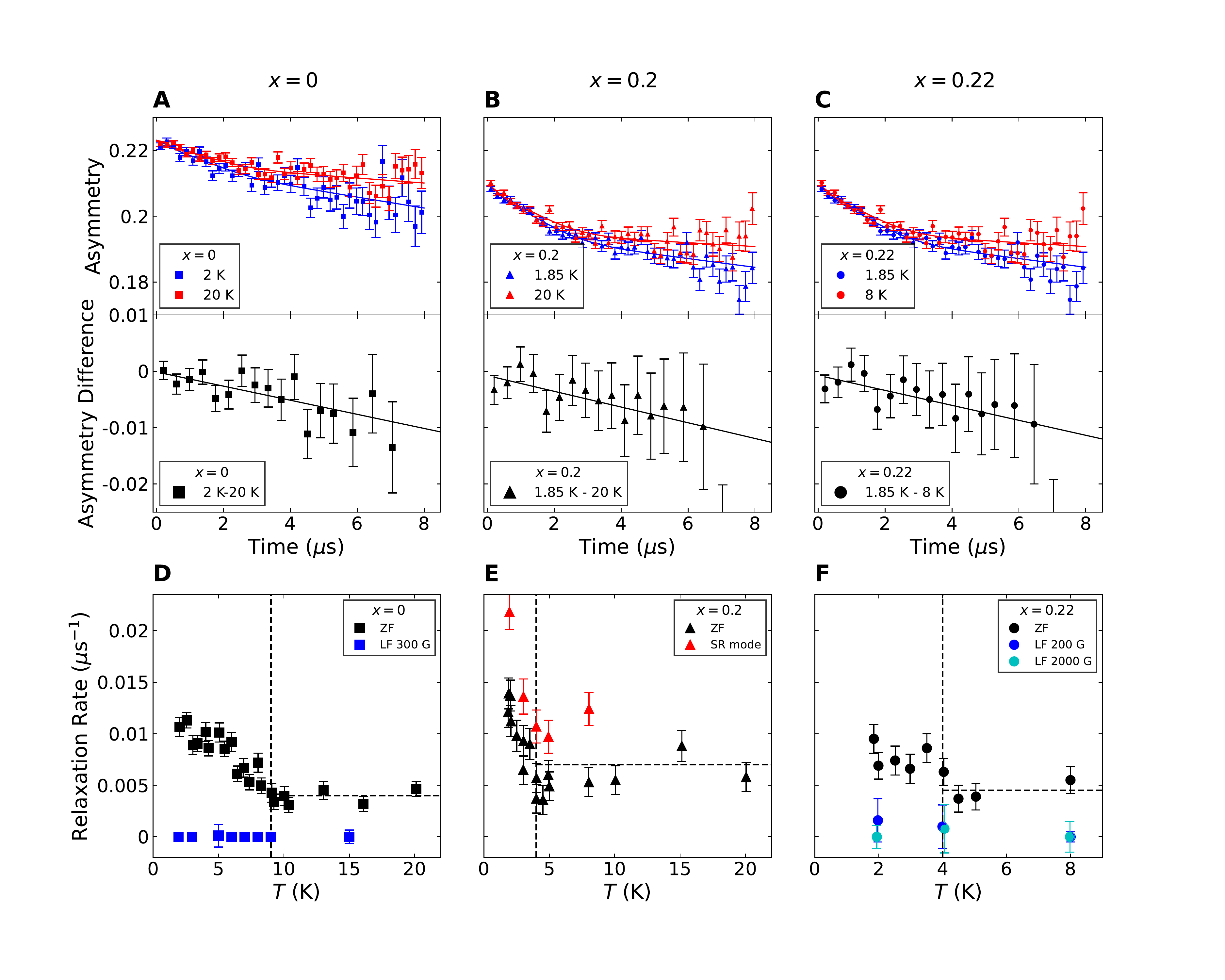}
	\vspace{-15mm}
	\caption{Evidence for TRS breaking superconductivity from zero-field $\bm{\mu}$SR. (A to C) Time evolution of $\mu$SR asymmetry in the NSR mode for FeSe (A) in the nematic phase, and for FeSe$_{1-x}$S$_{x}$ with $x=0.20$ (B) and 0.22 (C) in the tetragonal phase. Upper panels show the data compared at two temperatures above $T_{\rm c}$ in the normal state (red) and below $T_{\rm c}$ in the superconducting state (blue). Curves are fits to the relaxation curve described in Eq.\,(S1) in Supplementary Information. Lower panels show the difference of the data (symbols) between the two temperatures. The lines are the guides to the eyes. (D to F) Temperature dependence of the relaxation rate $\Lambda$ extracted from the fitting analysis for FeSe (D) in the nematic phase, and for FeSe$_{1-x}$S$_{x}$ with $x=0.20$ (E) and 0.22 (F) in the tetragonal phase. For all samples, we use the NSR mode (black), but for $x=0.20$ the SR mode is also used (red). The muon polarization is along the $c$ axis ($ab$-plane) in the NSR (SR) mode. In both modes, similar temperature dependence of $\Lambda$ is observed. For comparison, the results of relaxation rate in longitudinal-field ${\mu}$SR measurements up to 2000\,G (blue) are shown for $x=0$ (D) and $0.22$ (F). Vertical dashed lines mark the transition temperature $T_{\rm c}$. Horizontal dashed lines are guides to the eyes.
		}
	\label{fig:ZF}
\end{figure*}

TRSB in the superconducting state can be studied by the zero-field $\mu$SR technique, which is a sensitive probe of the internal magnetic field inside the sample~\cite{Luke1998}. When TRS is broken in the superconducting state, a small magnetic field can be induced near defects or TRSB domain boundaries~\cite{Matsumoto1999}, which can be detected by the change in the relaxation rate of $\mu$SR asymmetry. As shown in Fig.\:\ref{fig:ZF}\FigCap{a}-\FigCap{c}, we have measured zero-field $\mu$SR spectra at low temperatures for the collection of single crystals of FeSe ($x=0$) in the nematic phase, and FeSe$_{1-x}$S$_{x}$ ($x=0.20$ and 0.22) in the tetragonal phase. The comparisons of relaxation curves in the normal state above $T_{\rm c}$ (red symbols) and in the superconducting state below $T_{\rm c}$ (blue symbols) reveal a faster relaxation in the superconducting state for all samples. The relaxation rate $\Lambda$ of the samples can be extracted through the fitting analysis considering background contributions (see Supplementary Information), and the obtained temperature dependence of $\Lambda$ is shown in Fig.\:\ref{fig:ZF}\FigCap{d}-\FigCap{f}. The small and almost temperature-independent relaxation in the normal state is likely due to randomly oriented nuclear magnetic moments whose fluctuation speed is slower than the $\mu$SR time scale. For all the samples we measured, the relaxation rate increases below $T_{\rm c}$, and the increase in $\Lambda$ is observed in both non-spin-rotated (NSR) and spin-rotated (SR) configuration modes (Fig.\:\ref{fig:ZF}\FigCap{e}). These results indicate that a finite magnetic field develops in the sample in the superconducting state. Moreover, for the samples tested under longitudinal magnetic fields of 200--2000\,G ($x=0$ and 0.22), we find that the relaxation rate shows no enhancement below $T_{\rm c}$ (Fig.\:\ref{fig:ZF}\FigCap{d},\FigCap{f}), implying that the internal field below $T_{\rm c}$ observed in the zero-field $\mu$SR measurements is not fluctuating but static on the microsecond timescale. 
The increase of the relaxation rate is similar in magnitude both in the nematic and tetragonal phases, and the internal magnetic field at the lowest temperatures can be estimated as $B_{\rm int}\sim 0.09$, 0.12, and 0.07\,G for $x=0$, 0.20, and 0.22, respectively (Fig.\:\ref{fig:ZF}\FigCap{d}-\FigCap{f}). These values are significantly larger than the inevitable residual field ($<5$\,mG) in our shielded experimental setup. These results strongly indicate that TRS is broken in both nematic and tetragonal phases of superconducting FeSe$_{1-x}$S$_{x}$. 

In addition to the zero-field measurements, we have also performed the transverse-field $\mu$SR measurements under external fields of 330\,G. From the relaxation rate analysis in the superconducting state (see Supplementary Information), we can extract the temperature dependence of the superfluid density, which is proportional to $\lambda^{-2}\propto n_{\rm s}/m^*$, where $\lambda$ is the in-plane magnetic penetration depth and $m^*$ is the effective mass of quasiparticles. This gives information on the superconducting electron density $n_{\rm s}$, which can be compared with the total electron density $n$. The temperature dependence of $\lambda^{-2}$ for $x=0$ and 0.10 in the nematic phase and for $x=0.20$ and 0.22 in the tetragonal phase is shown in Fig.\:\ref{fig:superfluid}\FigCap{a}. The data for all samples deviate from the conventional $s$-wave curve, consistent with the anisotropic gap in this system. We also note that the $\lambda^{-2}(T)$ data for FeSe ($x=0$) are in good agreement with the previous $\mu$SR measurements for a 500-G $c$-axis field~\cite{Biswas2018} (Fig.\:S3). The absolute values of $\lambda^{-2}$ in the zero-temperature limit become smaller in the tetragonal phase compared with those in the nematic phase (Fig.\:\ref{fig:superfluid}\FigCap{a}, inset). Notably, this is opposite to the increasing trend of Fermi surface volumes as a function of $x$ in FeSe$_{1-x}$S$_{x}$ reported from the quantum oscillation measurements~\cite{Coldea2019}. 

More quantitatively, in Fig.\:\ref{fig:PD}\FigCap{c} we compare the $x$-dependence of $\lambda^{-2}(T\to 0)$ measured by $\mu$SR and that estimated from the Fermi surface structures based on quantum oscillations measurements~\cite{Coldea2019} (see Supplementary Information).  In FeSe ($x=0$), the $\lambda^{-2}(T\to 0)$ value is quantitatively consistent with the estimate from all measured orbits in quantum oscillations, indicating that all the carriers in the normal state are condensed into the superconducting pairing state. The comparison between measured and estimated $\lambda^{-2}$ in FeSe$_{1-x}$S$_{x}$ (Fig.\:\ref{fig:PD}\FigCap{c}) demonstrates that the superfluid density in the tetragonal phase is substantially suppressed from that expected from the normal-state electronic structure. Together with the reported phase diagram (Fig.\:\ref{fig:PD}\FigCap{a}) and the evolution of low-energy excitations with S composition $x$ (Fig.\:\ref{fig:PD}\FigCap{b}), we can summarize the unusual transformation of the superconducting state from SC1 phase to SC2 phase, driven by the disappearance of the nematic order in FeSe$_{1-x}$S$_{x}$, as follows. (i) The superconducting transition temperature $T_{\rm c}$ shows an abrupt change between SC1 in the nematic phase and SC2 in the tetragonal phase~\cite{Mizukami2021}. (ii) The zero-bias tunneling conductance~\cite{Hanaguri2018} and electronic specific heat~\cite{Mizukami2021,Sato2018} at low temperatures are both largely enhanced in the SC2 phase, indicating the presence of low-lying quasiparticle excitations in the tetragonal phase. (iii) The superfluid density shows a sizable suppression in the SC2 phase from that expected from the normal-state electronic structure. (iv) TRS is broken in both SC1 and SC2 phases. These results indicate that FeSe$_{1-x}$S$_{x}$ exhibits two different TRSB superconducting states SC1 and SC2, which are separated by the end point of the nematic order. The suppressed superfluid density in the tetragonal phase implies that not all electrons are condensed in the superconducting ground state, which is quite anomalous. The presence of unpaired electrons in the zero-temperature limit is not expected even in nodal superconductors, but consistent with the presence of zero-energy excitations reported by the STM and specific heat measurements. These results in the SC2 phase are consistent with an ultranodal pair state with emergent BFSs suggested theoretically.

\section*{Discussion}
Here we emphasize that the crystals used in this study are grown by the chemical vapor transport technique~\cite{Hosoi2016,Matsuura2017}, which is known as a method to obtain high-quality FeSe-based samples~\cite{Shibauchi2020}. The observation of quantum oscillations in a wide range indicates that the crystals are clean, and the STM results show that S atoms are distributed quite uniformly~\cite{Coldea2019,Hanaguri2018}. It is therefore highly unlikely that the low-lying excitations in SC2 originate from impurities or chemical inhomogeneities. This is reinforced by the STM measurements in the tetragonal phase, which have shown that the presence of the zero-bias conductance is a robust feature insensitive to the positions~\cite{Hanaguri2018}.
The fact that the onsets of these extraordinary excitations and the suppressions of $T_{\rm c}$ and $\lambda^{-2}$ match the fate of the nematic order strongly suggests that the emergence of ultranodal state is an intrinsic property triggered by the electronic change associated with the nematicity in this system.   

\begin{figure*}
	\centering
	\vspace{-25mm}
	\includegraphics[width=0.7\textwidth]{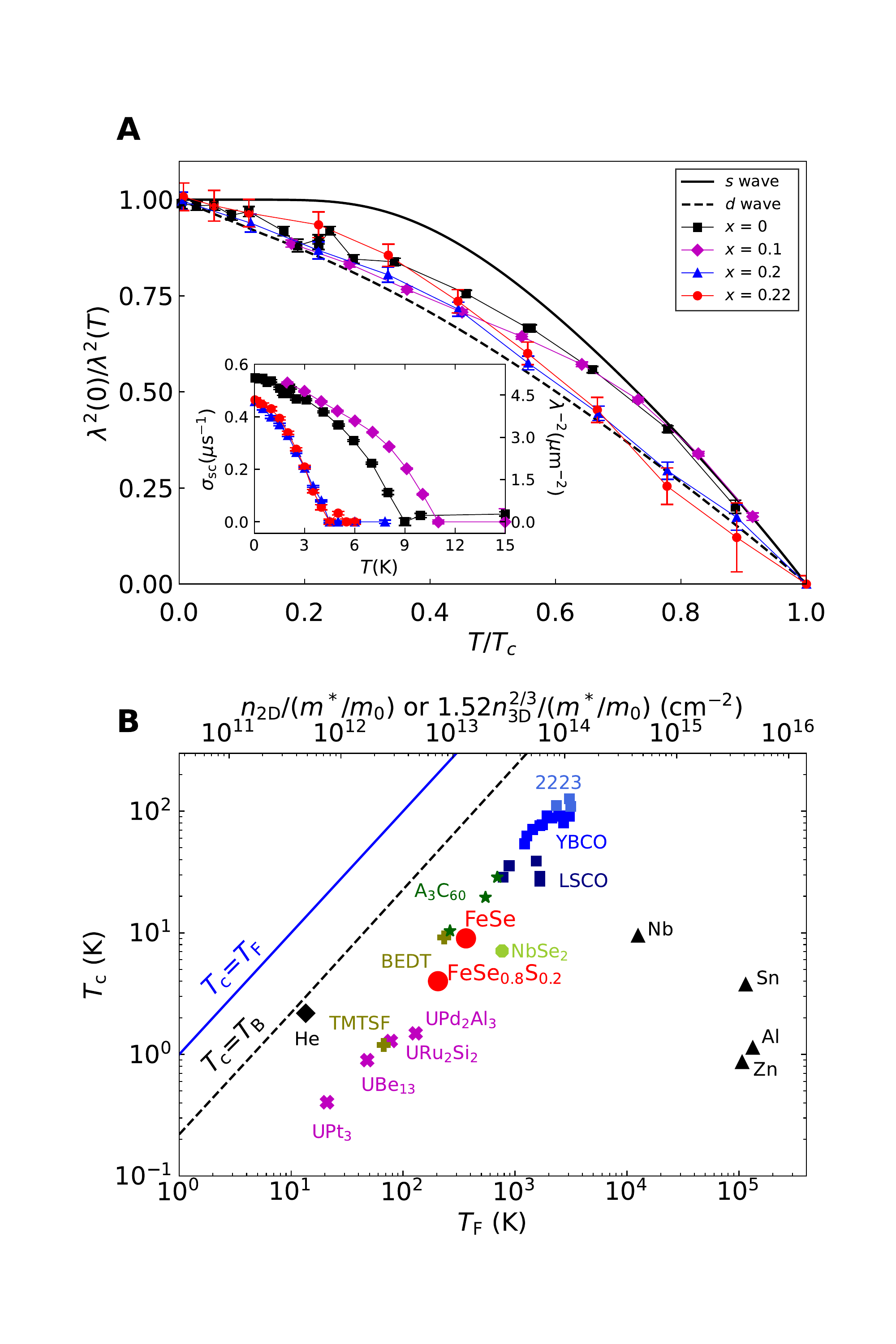}
	\vspace{-15mm}
	\caption{Superfluid density from transverse-field $\bm{\mu}$SR. (A) Normalized superfluid density $\lambda^2(0)/\lambda^2(T)$ as a function of reduced temperature $T/T_{\rm c}$ for FeSe$_{1-x}$S$_{x}$ with $x=0$, 0.10, 0.20, and 0.22. Solid (dashed) curve is the theoretical temperature dependence of superfluid density in the conventional fully gapped $s$-wave (line-nodal $d$-wave) superconductors. Inset shows the temperature dependence of relaxation rate $\sigma_{\rm sc}$ in the superconducting state (left axis) and superfluid density $\lambda^{-2}(T)$ without normalization (right axis). 
		(B) Uemura plot, where $T_{\rm c}$ is plotted against the effective superfluid density (upper axis). 
		The effective superfluid density is given by $n_{\rm 2D}/(m^*/m_0)$ and 1.52$n_{\rm 3D}$ for the 2D and 3D systems, respectively, where $n_{\rm 2D}$ is the carrier concentration in the superconducting plane of the 2D system, $n_{\rm 3D}$ is the carrier concentration of the 3D system, and $m_{0}$ is the mass of the free electron. 
		The Fermi temperature $T_{\rm F}$ (lower axis) is proportional to the effective carrier density $n_{\rm 2D}$, with the relation $T_{\rm F} = \hbar^{2}\pi n_{\rm 2D}/(k_Bm^*)$. In this study, we assumed that FeSe$_{1-x}$S$_x$ is in the limit of highly anisotropic superconductor and used a two-dimensional expression.
		The black dashed line is the Bose-Einstein condensation temperature of an ideal three-dimensional boson gas. 
		The solid blue line shows the line where $T_{\rm c} = T_{\rm F}$.	
	}
	\label{fig:superfluid}
\end{figure*}

One may also consider the effect of multigap, and the possibility of gapless Fermi surface in some bands, which could lead to the unpaired electrons. However, it has been widely discussed that the interband interactions are generally important for iron-based superconductors, and thus there must be nonnegligible coupling between different bands. In such a case, it is natural to consider that all the bands have superconducting ground states even if they have different gap magnitudes. In principle, some of the bands may have very small gaps, but the temperature dependence of the superfluid density down to very low temperatures (Fig.\:\ref{fig:superfluid}\FigCap{a}) does not show any signatures of such an extremely small gap. Thus we conclude that the multigap effect is unlikely the origin of the unpaired electrons.

We should note that the theoretical model of ultranodal states for FeSe$_{1-x}$S$_{x}$ does not provide microscopic origins at present~\cite{Setty2020}. Moreover, it is not clear how the BFSs can relate to nematicity. Thus, a complete understanding of the observed behaviors requires further studies. However, in recent theoretical calculations of the diamagnetic current response, it has been shown that there is a limited range of degree of TRSB to have stable BFSs, in which the superfluid density becomes less than half of the original value~\cite{Setty2020a}. This appears to be in good correspondence with the case for SC2 in the tetragonal phase of FeSe$_{1-x}$S$_{x}$.

The observed TRSB in the nematic phase of FeSe, in which no evidence of BFSs is found, also poses an intriguing question about the structure of the phase of the order parameter in the presence of nematic twin boundary. As mentioned above, near the nematic twin boundary, we expect a large change in the phase $\theta$ in the complex order parameter $\Delta_s+e^{{\rm i}\theta}\Delta_d$. If the TRS is preserved deep inside the bulk, we expect $\theta=0$ or $\pi$ away from the boundaries. Our zero-field $\mu$SR results show a similar change of relaxation rate below $T_{\rm c}$ in FeSe and tetragonal FeSe$_{1-x}$S$_{x}$ with and without nematic twin boundaries, which imply that TRS is broken inside the bulk even for FeSe. In such a nematic TRSB superconducting state, $\theta$ is shifted from 0 or $\pi$ in the bulk, but still, $\theta$ is different in the neighboring domains, and the phase can change near twin boundaries. This suggests that the degree of TRSB depends on the position near twin boundaries, which can explain the STM and ARPES results. Indeed, the original theory considering the orthorhombic domains by Sigrist {\it et al.}~\cite{Sigrist1996} has shown that the phase with broken TRS only near twin boundaries exists in a limited temperature region above the bulk TRSB state. In FeSe, the superconducting gap structure is very anisotropic and two-fold symmetric around the hole and electron bands~\cite{Sprau2017,Hashimoto2018}, which suggests that $\Delta_s$ and $\Delta_d$ have similar magnitudes. In such a case, it is plausible that the TRSB sets in very close to $T_{\rm c}$, which seems consistent with the observed enhancement of relaxation rate just below $T_{\rm c}$. 

\begin{figure*}
	\centering
	\includegraphics[width=0.6\textwidth]{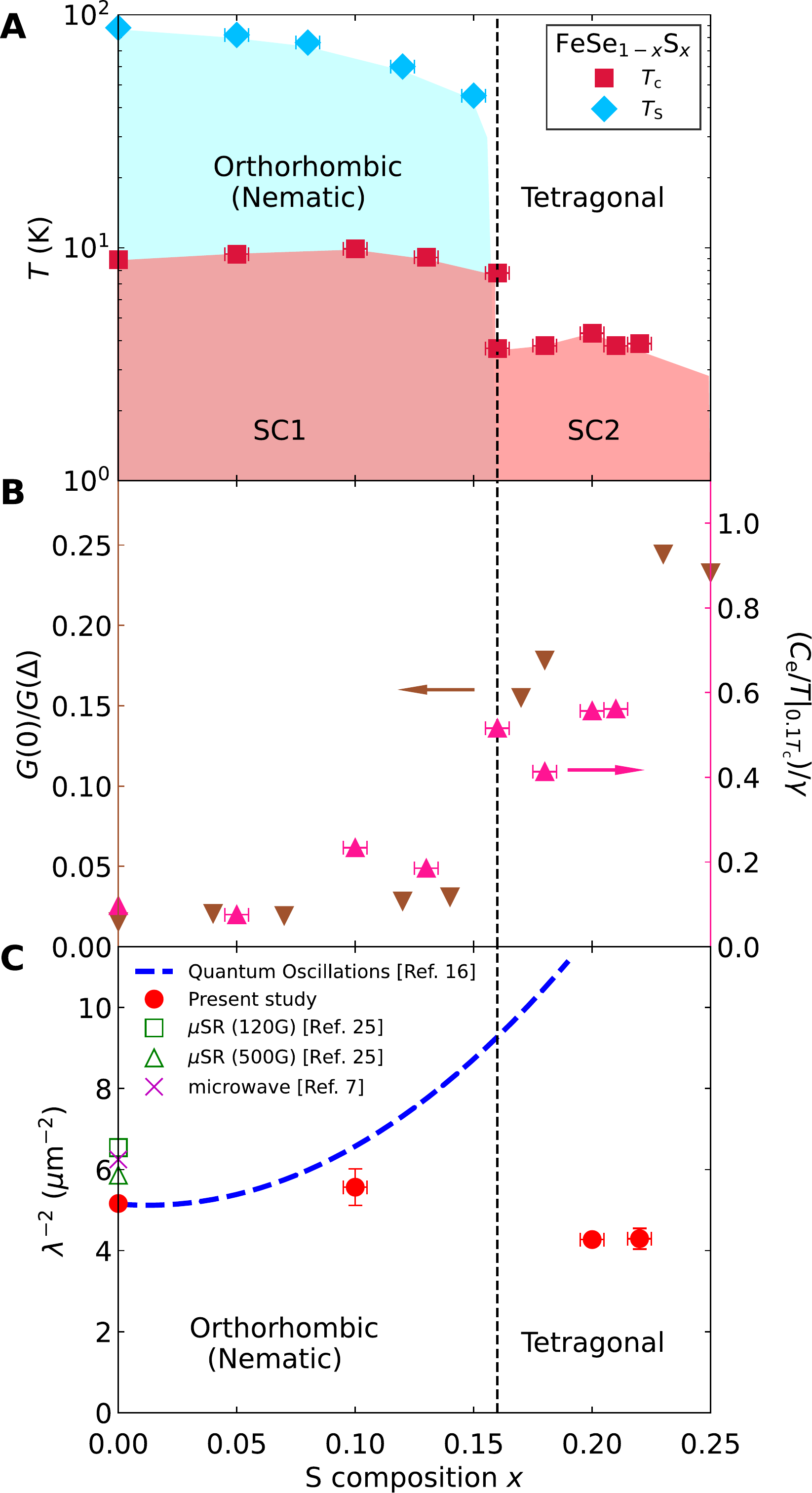}
	\caption{Phase diagram, low-energy excitations, and superfluid density in FeSe$\bm{_{1-x}}$S$\bm{_x}$. (A) Thermodynamically determined phase diagram of FeSe$_{1-x}$S$_{x}$. The high-temperature tetragonal to low-temperature orthorhombic (nematic) transition temperature $T_{\rm s}$ and superconducting transition temperature $T_{\rm c}$ are determined by specific heat measurements using vapor-grown single crystals~\cite{Mizukami2021}. (B) Zero-bias tunneling conductance $G(0)$ normalized by the conductance at the gap edge $G(\Delta)$ measured by STM~\cite{Hanaguri2018} and the low-temperature electronic specific heat divided by temperature $C_e/T$ at $T=0.1T_{\rm c}$ normalized by the Sommerfeld constant $\gamma$~\cite{Mizukami2021}, plotted against S composition $x$ in FeSe$_{1-x}$S$_{x}$ single crystals.  (C) Evolution of superfluid density in the zero-temperature limit (red circles) with $x$, compared with the estimated $\lambda^{-2}$ from the normal-state Fermi surface structure (dashed line) based on the quantum oscillations measurements~\cite{Coldea2019} (see Supporting Information). The previously reported values of $\lambda^{-2}$ in FeSe ($x=0$) from $\mu$SR measurements for 120\,G (cross) and 500\,G (square)~\cite{Biswas2018}, and from microwave surface impedance measurements in the Meissner state (triangle)~\cite{Kasahara2014} are also shown for comparison (see also Fig.\,S3). The vertical error bars are estimated from various fittings of $\lambda^{-2}(T)$. 
	}
	\label{fig:PD}
\end{figure*}

In addition to the TRSB and gap node topology, our present results for the superfluid density provide new insights on phenomenologies of unconventional superconductors. Figure\:\ref{fig:superfluid}\FigCap{b} shows correlations between $T_{\rm c}$ and the effective Fermi temperature $T_{\rm F} = E_{\rm F}$/$k_{\rm B}$ derived from the effective superfluid density $n_{\rm s}/m^*$~\cite{Uemura1991,Uemura2004}. The points from FeSe$_{1-x}$S$_{x}$ lie close to the nearly linear trend between $T_{\rm c}$ and $T_{\rm F}$ reported in hole-doped and electron-doped cuprates, organic, and heavy-fermion superconductors~\cite{Uemura1991,Uemura2004}, which is a hallmark of strongly correlated superconductivity. 
This may suggest that the sudden decrease in $T_{\rm c}$ above the nematic end point is closely related to the reduction in the superfluid density due to the emergence of BFSs in the tetragonal phase (see Fig.\:\ref{fig:PD}\FigCap{a},\FigCap{c}). 
The doping evolution of FeSe$_{1-x}$S$_{x}$ resembles the behaviors of overdoped cuprates, in the departure of superfluid density from the normal-state carrier density~\cite{Uemura2004}, existence of nematic quantum critical point~\cite{Ishida2020}, and orthorhombic to tetragonal transition. This analogy encourages high-precision experimental searches for TRSB in overdoped cuprates.

Our $\mu$SR measurements reveal two different TRSB superconducting states in the phase diagram of FeSe$_{1-x}$S$_{x}$. The suppression of nematicity leads to the reduction in $T_{\rm c}$ as well as the suppressed superfluid density, which indicates that the superconductivity in the tetragonal phase is very exotic, having broken TRS and unpaired electrons. The results are consistent with the ultranodal state with BFSs proposed theoretically, but details of the BFSs including the size and the location in the momentum space require further studies. 
Together with the recent reports of Bose-Einstein condensation-like superconductivity in FeSe$_{1-x}$S$_{x}$~\cite{Hashimoto2020,Mizukami2021}, our findings may open up a new field of studies on the extended zero-energy excitations in fermionic bound states.

\matmethods{{\bf Single Crystals.}
Single crystals of FeSe$_{1-x}$S$_{x}$ ($x = 0, 0.10, 0.20,$ and 0.22) used in this study were synthesized by the chemical vapor transport method \cite{Hosoi2016,Matsuura2017}. For the $\mu$SR measurements, a sample mass density of about 200\,mg/cm$^2$ is required to stop the muon in the sample. 
For the pure FeSe samples, typical crystal dimensions are about $1\times1$\,mm$^2$ in the $ab$ plane and several tens of $\mu$m in the $c$-axis direction, and weighs about 1 to 5\,mg. 
The sizes of the S-substituted samples are even smaller. 
Therefore, for each $x$, a large number of single-crystal samples (up to 100 pieces) were co-aligned on a silver plate with Apiezon N grease (whose volume is around 10\% of the sample mass) to cover an area of $\sim1$\,cm$^2$, so that the $c$ axes were aligned perpendicular to the plate. 
For $x = 0$, $0.10$ and $0.22$, crystals were collected from a single batch each, but for $x=0.20$ we used a few batches to extract enough crystals with a total mass of $\sim 170$\,mg. 
Energy-dispersive X-ray spectroscopy was used to determine the compositions of S-substituted samples by averaging the $x$ values obtained for several samples taken from the batches. 
For $x=0.22$, we also measured the temperature dependence of resistivity on a crystal taken from the same batch, and it was confirmed that there was no anomaly due to structural phase transition and that the superconducting transition temperature was around 4\,K. 
In addition, the $c$-axis lattice parameters determined by X-ray diffraction experiments are consistent with the previous reports \cite{Hosoi2016,Matsuura2017}.
\\
{\bf $\mu$SR measurements analysis.} 
$\mu$SR measurements were performed at the Centre for Molecular and Materials Science at TRIUMF in Vancouver, Canada. We conducted zero-field $\mu$SR measurements above 2\,K with the M20D beamline, while the measurements below 2\,K were conducted on the M15 surface muon channel with a dilution refrigerator. Stray magnetic fields at the sample position at M20D were measured by fluxgate magnetometer and found to be $\sim 2$, 1, and 5\,mG along 3 orthogonal directions. Stray fields were reduced to less than 5\,mG in all directions in the dilution refrigerator using the method of Ref.\,\cite{Morris2003}. The $\mu$SR analysis was performed in the time domain using the program MUSRFIT~\cite{Suter2012}. More detailed information for the procedure of data analysis is described in Supporting Information.
\\\\
{\bf Data Availability.} All data are included in the manuscript and Supporting Information.
}

\showmatmethods{} 

\acknow{We thank D.\ F.~Agterberg, P.\ M.\ R.~Brydon, R.\ M.~Fernandes, T.~Hanaguri, P.\ J.~Hirschfeld, K.~Kuboki, Y.~Matsuda, E.-G.~Moon, C.~Setty, and M.~Sigrist for fruitful discussions. 
	{\bf Funding:} This work was supported by Grants-in-Aid for Scientific Research (KAKENHI) (Nos.\ JP22H00105, JP21H01793, JP19H00649, JP18H05227, JP18KK0375), Grant-in-Aid for Scientific Research on innovative areas ``Quantum Liquid Crystals" (No.\ JP19H05824) and Grant-in-Aid for Scientific Research for Transformative Research Areas (A) “Condensed Conjugation” (No.\ JP20H05869) from Japan Society for the Promotion of Science (JSPS), and CREST (No.\ JPMJCR19T5) from Japan Science and Technology (JST). The work at Columbia and TRIUMF has been supported by the US National Science Foundation Grant No.\ DMR-1610633 and the DMREF Project No.\ DMR-1436095, the Reimei Project from the Japan Atomic Energy Agency (JAEA), and a support from the Friends of Tokyo University Inc.
	G.Q.Z. has been supported in part by China Scholarship Council (No.\ 201904910900).
	This research was undertaken thanks in part to funding from the Max Planck-UBC-UTokyo Centre for Quantum Materials and the Canada First Research Excellence Fund, Quantum Materials and Future Technologies Program.
}

\showacknow{} 

\bibliography{ref_TRSB}

\begin{thebibliography}{10}

\bibitem{Shibauchi2020}
Shibauchi T, Hanaguri T, Matsuda Y (2020) Exotic superconducting states in
  fese-based materials.
\newblock {\em J. Phys. Soc. Jpn.} 89(10):102002.

\bibitem{Coldea2021}
Coldea AI (2021) {Electronic nematic states tuned by isoelectronic substitution
  in bulk FeSe$_{1-x}$S$_x$}.
\newblock {\em Front. Phys.} 8:594500.

\bibitem{Hosoi2016}
Hosoi S, et~al. (2016) Nematic quantum critical point without magnetism in
  {FeSe}$_{1-x}${S}$_x$ superconductors.
\newblock {\em Proc. Natl. Acad. Sci. USA} 113(29):8139--8143.

\bibitem{Ishida2022}
Ishida K, et~al. (2022) Pure nematic quantum critical point accompanied by a
  superconducting dome.
\newblock {\em Proc. Natl. Acad. Sci. USA} 119(18):e2110501119.

\bibitem{Licciardello2019}
Licciardello S, et~al. (2019) Electrical resistivity across a nematic quantum
  critical point.
\newblock {\em Nature} 567(7747):213--217.

\bibitem{Huang2020}
Huang WK, et~al. (2020) {Non-Fermi} liquid transport in the vicinity of the
  nematic quantum critical point of superconducting {FeSe$_{1-x}$S$_{x}$}.
\newblock {\em Phys. Rev. Research} 2(3):033367.

\bibitem{Kasahara2014}
Kasahara S, et~al. (2014) {Field-induced superconducting phase of FeSe in the
  BCS-BEC cross-over}.
\newblock {\em Proc. Natl. Acad. Sci. USA} 111(46):16309--16313.

\bibitem{Watashige2015}
Watashige T, et~al. (2015) Evidence for time-reversal symmetry breaking of the
  superconducting state near twin-boundary interfaces in fese revealed by
  scanning tunneling spectroscopy.
\newblock {\em Phys. Rev. X} 5(3):031022.

\bibitem{Xu2016}
Xu HC, et~al. (2016) Highly anisotropic and twofold symmetric superconducting
  gap in nematically ordered ${\mathrm{fese}}_{0.93}{\mathrm{s}}_{0.07}$.
\newblock {\em Phys. Rev. Lett.} 117(15):157003.

\bibitem{Sprau2017}
Sprau PO, et~al. (2017) Discovery of orbital-selective cooper pairing in
  {FeSe}.
\newblock {\em Science} 357(6346):75--80.

\bibitem{Hashimoto2018}
Hashimoto T, et~al. (2018) Superconducting gap anisotropy sensitive to nematic
  domains in {FeSe}.
\newblock {\em Nat. Commun.} 9(1):282.

\bibitem{Liu2018}
Liu D, et~al. (2018) Orbital origin of extremely anisotropic superconducting
  gap in nematic phase of fese superconductor.
\newblock {\em Phys. Rev. X} 8(3):031033.

\bibitem{Sigrist1996}
Sigrist M, Kuboki K, Lee PA, Millis AJ, Rice TM (1996) Influence of twin
  boundaries on josephson junctions between high-temperature and conventional
  superconductors.
\newblock {\em Phys. Rev. B} 53(5):2835--2849.

\bibitem{Kang2018}
Kang J, Chubukov AV, Fernandes RM (2018) Time-reversal symmetry-breaking
  nematic superconductivity in {FeSe}.
\newblock {\em Phys. Rev. B} 98(6):064508.

\bibitem{Hanaguri2018}
Hanaguri T, et~al. (2018) Two distinct superconducting pairing states divided
  by the nematic end point in {FeSe}$_{1-x}${S}$_x$.
\newblock {\em Sci. Adv.} 4(5):eaar6419.

\bibitem{Coldea2019}
Coldea AI, et~al. (2019) Evolution of the low-temperature {Fermi} surface of
  superconducting {FeSe}$_{1-x}${S}$_x$ across a nematic phase transition.
\newblock {\em npj Quant. Mater.} 4(1):2.

\bibitem{Sato2018}
Sato Y, et~al. (2018) Abrupt change of the superconducting gap structure at the
  nematic critical point in {FeSe}$_{1-x}${S}$_x$.
\newblock {\em Proc. Natl. Acad. Sci. USA} 115(6):1227--1231.

\bibitem{Mizukami2021}
Mizukami Y, et~al. (2021) {BCS-BEC crossover superconductivity in
  FeSe$_{1-x}$S$_x$: Thermodynamics of the transition}.
\newblock {\em preprint} p. arXiv:2105.00739.

\bibitem{Agterberg2017}
Agterberg DF, Brydon PMR, Timm C (2017) Bogoliubov {Fermi} surfaces in
  superconductors with broken time-reversal symmetry.
\newblock {\em Phys. Rev. Lett.} 118(12):127001.

\bibitem{Brydon2018}
Brydon PMR, Agterberg DF, Menke H, Timm C (2018) {Bogoliubov Fermi surfaces:
  General theory, magnetic order, and topology}.
\newblock {\em Phys. Rev. B} 98(22):224509.

\bibitem{Setty2020}
Setty C, Bhattacharyya S, Cao Y, Kreisel A, Hirschfeld PJ (2020) Topological
  ultranodal pair states in iron-based superconductors.
\newblock {\em Nat. Commun.} 11(1):523.

\bibitem{Setty2020a}
Setty C, Cao Y, Kreisel A, Bhattacharyya S, Hirschfeld PJ (2020) {Bogoliubov
  Fermi surfaces in spin-$\frac{1}{2}$ systems: Model Hamiltonians and
  experimental consequences}.
\newblock {\em Phys. Rev. B} 102(6):064504.

\bibitem{Luke1998}
Luke GM, et~al. (1998) Time-reversal symmetry-breaking superconductivity in
  {Sr}$_2${RuO}$_4$.
\newblock {\em Nature} 394(6693):558--561.

\bibitem{Matsumoto1999}
Matsumoto M, Sigrist M (1999) Quasiparticle states near the surface and the
  domain wall in a $p_x\pm {\rm i}p_y$-wave superconductor.
\newblock {\em J. Phys. Soc. Jpn.} 68(3):994--1007.

\bibitem{Biswas2018}
Biswas PK, et~al. (2018) Evidence of nodal gap structure in the basal plane of
  the {FeSe} superconductor.
\newblock {\em Phys. Rev. B} 98(18):180501.

\bibitem{Matsuura2017}
Matsuura K, et~al. (2017) Maximizing {$T_c$} by tuning nematicity and magnetism
  in {FeSe}$_{1-x}${S}$_x$ superconductors.
\newblock {\em Nat. Commun.} 8(1):1143.

\bibitem{Uemura1991}
Uemura YJ, et~al. (1991) {Basic similarities among cuprate, bismuthate,
  organic, Chevrel-phase, and heavy-fermion superconductors shown by
  penetration-depth measurements}.
\newblock {\em Phys. Rev. Lett.} 66(20):2665--2668.

\bibitem{Uemura2004}
Uemura YJ (2004) {Condensation, excitation, pairing, and superfluid density in
  high-$_{\rm c}$ superconductors: the magnetic resonance mode as a roton
  analogue and a possible spin-mediated pairing}.
\newblock {\em J. Phys.: Condens. Matter} 16(40):S4515.

\bibitem{Ishida2020}
Ishida K, et~al. (2020) Divergent nematic susceptibility near the pseudogap
  critical point in a cuprate superconductor.
\newblock {\em J. Phys. Soc. Jpn.} 89(6):064707.

\bibitem{Hashimoto2020}
Hashimoto T, et~al. (2020) {Bose-Einstein condensation superconductivity
  induced by disappearance of the nematic state}.
\newblock {\em Sci. Adv.} 6(45):eabb9052.

\bibitem{Morris2003}
Morris GD, Heffner RH (2003) A method of achieving accurate zero-field
  conditions using muonium.
\newblock {\em Physica B} 326:252--254.

\bibitem{Suter2012}
Suter A, Wojek B (2012) {Musrfit: A} free platform-independent framework for
  {$\mu$SR} data analysis.
\newblock {\em Phys. Proc.} 30:69--73.
\newblock 12th International Conference on Muon Spin Rotation, Relaxation and
  Resonance ($\mu$SR2011).

\end{thebibliography}

\clearpage


\end{document}